\documentclass[useAMS,usenatbib,twocolumn]{mnras}
\usepackage{amsmath,slashed}

\usepackage{color}

\usepackage{comment}
\usepackage{graphicx}

\usepackage{amsmath,amssymb}
\usepackage{bm}
\usepackage{enumitem}

\usepackage{natbib}
\usepackage{wasysym}

\renewcommand{\em}{{\rm em}}
\newcommand{\emO}{{\rm em,0}}
\newcommand{\gw}{{\rm gw}}

\renewcommand{\sf}{{\rm sf}}

\newcommand{\be}{\begin{equation}}
\newcommand{\ee}{\end{equation}}
\newcommand{\bear}{\begin{eqnarray}}
\newcommand{\eear}{\end{eqnarray}}

\newcommand{\rp}{{\rm p}}

\title[Gravitational waves from short GRBs]{Observationally constraining gravitational wave emission from short gamma-ray burst remnants}
\author[P. D. Lasky \& K. Glampedakis]{Paul D. Lasky$^1$\thanks{paul.lasky@monash.edu} and Kostas Glampedakis$^{2}$\thanks{kostas@um.es}\\
$^1$Monash Centre for Astrophysics, School of Physics and Astronomy, Monash University, VIC 3800, Australia\\
$^2$Departamento de F\'i'sica, Universidad de Murcia, Murcia, E-30100, Spain\\
}

\begin{document}

\pagerange{\pageref{firstpage}--\pageref{lastpage}} \pubyear{?}

\maketitle

\label{firstpage}

\begin{abstract}
Observations of short gamma-ray bursts indicate ongoing energy injection following the prompt emission, with the most likely candidate 
being the birth of a rapidly rotating, highly magnetised neutron star.  We utilise X-ray observations of the burst remnant to constrain properties 
of the nascent neutron star, including its magnetic field-induced ellipticity and the saturation amplitude of various oscillation modes.  
Moreover, we derive strict upper limits on the gravitational wave emission from these objects by looking only at the X-ray light curve, showing the burst remnants are unlikely to be detected in the near future using ground-based gravitational wave interferometers such as Advanced LIGO.  
\end{abstract}


\maketitle


\section{Introduction}
\label{sec:intro}
Recent observations of short and long gamma-ray bursts (GRBs) provide evidence for ongoing energy injection in their central core, exhibited as extended (lasting $\sim10^3$ -- $10^6$ s) X-ray emission.  Debate rages over the exact nature of the central engine, be they hyper-accreting black holes \citep[e.g.,][]{popham99,kisaka15}, or rapidly spinning, strongly magnetised neutron stars \citep[e.g.,][]{dai98a,dai98b,zhang01}.  Extended emission from {\it short} GRBs is consistent with the millisecond magnetar model \citep{rowlinson13,rowlinson14,lu15}, with some authors suggesting gravitational waves from the highly-deformed neutron star remnant may be detectable by ground-based gravitational wave interferometers \citep{fan13,dallosso15,doneva15}.  In this paper, we derive strict, {\it empirical} upper limits on the gravitational wave emission from neutron star remnants driving X-ray plateaus following short GRBs.

The progenitors of short GRBs are thought to be mergers of binary compact objects \citep[neutron stars and/or black holes; see][for a comprehensive review]{berger14}.  For the millisecond magnetar model, the subset of short GRBs with extended emission can only be produced by binary neutron star mergers.  Recent observations of old, massive neutron stars \citep[$\sim2\,M_\odot$;][]{demorest10,antoniadis13}, and our lack of understanding of the neutron star equation of state, allow for the theoretical possibility that binary neutron star mergers can result in massive neutron star remnants \citep[e.g.,][]{zhang13a,giacomazzo13,lasky14,ravi14b}.

Observations of short GRBs using the {\it Swift} satellite show a plateau phase lasting $t_b\sim10$ -- 100 s, before a decay phase where the luminosity 
is consistent with $L\propto t^{-2}$ \citep{rowlinson13}.  Such an evolution is the one exactly predicted by the millisecond magnetar model, where dipole 
radiation dominated spindown of the nascent neutron star drives the X-ray plateau and decay \citep[][]{zhang01,metzger14b}.  Adding weight to this model is the 
observation of a subset of these short GRBs that exhibit a steep decay phase at $t\gtrsim$100 s.  Within the millisecond magnetar model, this is interpreted as 
the collapse of the supramassive neutron star to a black hole.  

In this paper, we use properties of the X-ray light curves to place constraints on properties of the nascent neutron star.  In particular, we constrain the ellipticity of the 
neutron star and the total energy emitted in gravitational radiation.  We derive the limits on gravitational wave emission in a simple way: if gravitational wave 
emission {\it dominated} the spindown of the neutron star (over electromagnetic torques), then the luminosity evolution of the X-ray plateau would take on a different 
form to that of dipole radiation-dominated spindown \citep{zhang01}.  The fact that dipole radiation dominated spindown fits the evolution of the X-ray light curve so 
well \citep[e.g.,][]{rowlinson13} allows us to put strict upper limits on the amount of energy being lost to the system through gravitational wave emission (see Section \ref{theory} 
for details of the method).  In turn, depending on the mechanism generating those gravitational waves, we place constraints on various properties of the neutron star, 
including its ellipticity and the maximum (saturation) amplitude of neutron star oscillation modes (Sections \ref{sec:epsall} and \ref{ellipticity_constraints}). 

Numerous authors \citep[e.g.,][]{corsi09a,dallosso15,doneva15} have argued, on theoretical grounds, that gravitational wave emission from neutron stars born in binary mergers may be observable by second-generation gravitational wave interferometers such as Advanced LIGO \citep[aLIGO;][]{ligo15}.  However, while theoretically plausible, if gravitational wave energy loss were so significant, it would have a marked effect on the evolution of the X-ray light curve, which is inconsistent with observations.  We show that the prospects for gravitational wave detection from these systems in the next few years are bleak (see Section \ref{sec:GWcons}).  Upper limits on the total gravitational wave emission indicate that aLIGO is not likely to detect these systems, even when operating at full, design sensitivity.  Even the third-generation Einstein Telescope \citep{punturo10} has only a small chance of detection. This rather pessimistic prediction could of course fail for a short GRB event occurring nearby, say, at $\sim 40\,\mbox{Mpc}$ (i.e. the distance of the closest -- albeit long -- GRB event detected so far). 

The paper is set out as follows:  In Section \ref{theory}, we establish how gravitational wave emission can be constrained through observations of the X-ray plateaus following short GRBs.  In Section \ref{sec:epsall} we derive equations for the total, theoretical upper limit of the ellipticity (Section \ref{sec:epsobs}), and subsequently relate this to physical mechanisms active in the star, including magnetic field-induced deformations (Sections \ref{sec:epsb} and \ref{sec:epssf}), bar modes (Section \ref{sec:bar}) and inertial modes (Section \ref{sec:inertial}).  In Section \ref{ellipticity_constraints} we relate these theoretical arguments to a sample of eight, short GRBs to derive constraints on the nascent neutron stars in those systems.  Finally, in Section \ref{sec:GWcons} we derive constraints on the total gravitational wave emission from this sample, and show the detectability of their signals in first-, second- and third-generation gravitational wave interferometers.  

Some parts of Sections \ref{sec:epsall} and \ref{ellipticity_constraints} are technical with respect to physics of the newly-born neutron star.  However, a fast-track reader can obtain the main gravitational wave constraints through the simpler analysis presented in Section \ref{sec:GWcons}.


\section{Theoretical light curves}
\label{theory}

Consider a newly born neutron star being spun down through a combination of electromagnetic dipole and gravitational wave quadrupole emission.  
If gravitational wave emission comes from non-axisymmetries in the neutron star, the spin down law is \citep[e.g.,][]{shapiro83}
\begin{align}
	-I\Omega\dot{\Omega}=\frac{B_p^2R^6\Omega^4}{6c^3}+\frac{32GI^2\epsilon^2\Omega^6}{5c^5},\label{spindown}
\end{align}
where $\Omega$ and $\dot{\Omega}$ are the angular frequency and its time derivative, $I$ is the moment of inertia, $B_{p}$ is the dipole component of 
the magnetic field at the pole, $\epsilon$ is the ellipticity, $R$ is the neutron star radius, and $c$ and $G$ are the speed of light and Newton's gravitational 
constant, respectively. 

Although the evolution of $\Omega(t)$ includes both electromagnetic and gravitational wave contributions, the X-ray luminosity is simply given by the 
energy input into the surrounding medium from electromagnetic losses \citep{zhang01,metzger14b}.  That is
\begin{align}
	L(t)=\frac{\eta B_p^2 R^6\Omega(t)^4}{6c^3},\label{lem}
\end{align}
where $\Omega(t)$ is given by the full solution of equation (\ref{spindown}).  A fudge factor, $\eta\le1$, has been included in equation (\ref{lem}) that accounts for imperfect efficiency in converting spin-down energy into electromagnetic radiation, particularly through the X-ray channel.  \citet{rowlinson14} used the empirical correlation between the short GRB X-ray plateaus and their intrinsic luminosities to derive a relationship between the efficiency and the beaming angle.  For a beaming angle of $\sim$8--12$^{\rm o}$, the efficiency is $\eta\approx0.1$ \citep[see Figure 3 of][]{rowlinson14}.  As we show below, $\eta<1$ lowers the upper limit of gravitational wave emission inferred from the observations; setting $\eta=1$ therefore gives a true upper limit.  We retain $\eta$ throughout the derivation as it is an important parameter when one tries to constrain the physics of the nascent neutron star (e.g., the existence and behaviour of bar mode instabilities) from the results presented herein.

It is instructive to consider the cases when electromagnetic radiation or gravitational wave emission dominate spindown.  The former case assumes the final 
term in equation (\ref{spindown}) is negligibly small, implying the X-ray luminosity becomes
\begin{align}
	L(t)=L_\emO\left(1+\frac{t}{\tau_{\rm em}}\right)^{-2},\label{lemdom}
\end{align}
where 
\begin{align}
	L_\emO=\frac{\eta I\Omega_0^2}{2\tau_\em},\label{Lem0}
\end{align}
and $\Omega_0$ are the luminosity and angular frequency at $t=0$, respectively\footnote{Neutrino-driven winds initially accelerate the spin down of the nascent neutron star \citep{thompson04,metzger11}, implying the ``initial'' quantities we refer to are those determined when dipole and/or gravitational wave emission comes to dominate the spindown $\lesssim100$ s after the merger.  This timescale is similar to that with which {\it Swift} XRT begins collecting data, implying neutron star parameters inferred from $X$-ray plateaus \citep[e.g.,][]{rowlinson13} are also those following the neutrino-driven wind phase.}, and the electromagnetic spindown timescale is
\begin{align}
	\tau_\em=\frac{3c^3I}{B_p^2R^6\Omega_0^2}.
\end{align}
Equation (\ref{lemdom}) implies the light curve shows a distinct plateau feature, $L=L_\emO$ for $t\ll \tau_\em$, before decaying as $L=L_\emO(t/\tau_\em)^{-2}$ for $t\gg\tau_\em$.  This remarkably simple behaviour has been observed in the remnants of a number of short \citep[][]{rowlinson10a,rowlinson13,gompertz13,lu14,deugartepostigo14,lu15} and long \citep[][]{troja07,lyons10,dallosso11,bernardini12,yi14} GRB afterglows, which not only supports the magnetar model, but has been used to infer the initial spin period and magnetic field strength of remnants.  

Conservation of angular momentum implies the merger remnant should be rotating at, or near, break-up.  Such rapid rotations are conducive to large deformations and/or mode oscillation amplitudes, implying significant losses to gravitational radiation are possible.   Indeed, some have suggested gravitational wave emission could dominate spindown \citep{fan13,dallosso15}.  However, gravitational wave dominated spindown has a characteristic luminosity evolution different to dipole radiation dominated spindown \citep{zhang01}.  In particular, assuming the first term on the right hand side of (\ref{spindown}) is negligibly small implies the luminosity is given by
\begin{align}
	L(t)=L_\emO\left(1+\frac{t}{\tau_\gw}\right)^{-1},\label{lgwdom}
\end{align}
where the gravitational wave spindown timescale for an optimal emitter of gravitational waves (i.e., where the principal eigenvector of the moment of inertia tensor is orthogonal to the stellar rotation axis) is
\begin{align}
	\tau_\gw=\frac{5c^5}{128GI\epsilon^2\Omega_0^4}.\label{tauGW}
\end{align}
Equation (\ref{lgwdom}) shows that a system dominated by gravitational wave emission undergoes a plateau phase at early times, $L=L_\emO$ for $t\ll\tau_\gw$, but decays as $L=L_\emO(t/\tau_\gw)^{-1}$ for $t\gg\tau_\gw$.  Note that the characterirstic plateau timescale is now the gravitational wave, rather than the electromagnetic, spindown timescale, and the luminosity decay at late times goes as $\propto t^{-1}$ rather than $\propto t^{-2}$ for dipole spindown.

In reality, if gravitational wave emission is to dominate spindown, then it must do so at early times in the evolution (see below for proof).  We therefore define a transition timescale, $\tau_\star$, as the time at which point the spindown luminosity changes from being gravitational wave dominated to being dominated by elecgtromagnetic emission \citep{zhang01}.  Therefore, $\tau_\star$ is defined as the time that satisfies $L_\gw(\tau_\star)=L_\em(\tau_\star)$, where $L_\gw$ is the luminosity given by gravitational wave-only spindown, i.e., equation (\ref{lgwdom}), and $L_\em$ is the luminosity for only electromagnetic dominated spindown, equation (\ref{lemdom}).  Evaluating and rearranging gives 
\begin{align}
	\tau_\star=\frac{\tau_\em}{\tau_\gw}\left(\tau_\em-2\tau_\gw\right).\label{taustar}
\end{align}
Equation (\ref{taustar}) has two distinct regimes: (i) If $\tau_\gw>\tau_\em/2$, then $\tau_\star<0$, implying electromagnetic dipole emission always dominates spindown, and (ii) if $\tau_\gw<\tau_\em/2$, gravitational wave emission dominates spindown at early times, but electromagnetic emission dominates late.  

For any given observation that exhibits both a plateau phase followed by some decay, an upper limit on the gravitational wave emission can be attained by assuming case (ii) above.  That is, that gravitational wave emission dominates at early times.  If the decay of the lightcurve goes as $L\propto t^{-2}$, then the upper limit on the gravitational wave emission can be found by assuming gravitational wave emission dominates early, but electromagnetic emission takes over at $\tau_\star\le t_b$, where $t_b$ is the plateau timescale (i.e. the time marking the temporal ``break" in the light curve).


\section{Ellipticity constraints}
\label{sec:epsall}
Central to the calculation of gravitational wave emission from the nascent neutron star is the allowed range of ellipticities the star can have.  There are four different ellipticities that we can infer from the X-ray light curves, all of which are relevant.  Firstly, the maximal ellipticity based solely on the light curve as described in Section \ref{theory} (see Section \ref{sec:epsobs}).  Secondly, an expected ellipticity induced from magnetic field deformations (Section \ref{sec:epsb}).  Thirdly, a critical ellipticity, above which the star will remain axisymmetric despite the large deformation, and hence the star cannot emit substantial gravitational wave energy (Section \ref{sec:epssf}).  Finally, an effective ellipticity that represents excited oscillation modes such as $f$-modes (Section \ref{sec:bar}) and inertial modes (Section \ref{sec:inertial}).

\subsection{Observational limits on ellipticity}\label{sec:epsobs}
The sample of short GRBs is characterised by light cures consistent with $L_\em\propto t^{-2}$ for $t>t_b$, with the favoured energy-injection scenario being due to dipole radiation spindown of the nascent neutron star \citep[][]{rowlinson13,lu15}.  The transition timescale, $\tau_\star$, must necessarily occur before the break, otherwise the lightcurve following the break would decay as $t^{-1}$, rather than the observed $t^{-2}$ (see Section \ref{theory}).  Moreover, this indicates (as shown above) that the break time is the electromagnetic spindown timescale; i.e., $\tau_\star\le t_b=\tau_{\em}$.  Combining this condition with equation (\ref{taustar}) implies the gravitational wave spindown timescale is constrained to be $\tau_\gw \ge t_b/3$ which, when further combined with equations (\ref{Lem0}) and (\ref{tauGW}), gives
\begin{align}
	\epsilon_{\rm obs}&\le\left(\frac{15 c^5\eta^2 I}{512GL_\emO^2t_b^3}\right)^{1/2}\notag\\
	&=0.33\eta \left(\frac{I}{10^{45}\,{\rm g\,cm}^2}\right)^{1/2}\left(\frac{L_\emO}{10^{49}\,{\rm erg\,s}^{-1}}\right)^{-1}\left(\frac{t_b}{100\,{\rm s}}\right)^{-3/2}.\label{internaleps}
\end{align}
The inequality in (\ref{internaleps}) allows us to calculate an upper limit for the ellipticity given only observables (and the moment of inertia) associated with the X-ray lightcurve.  

It is worth stressing that $\epsilon_{\rm obs}$ represents the case of an optimal emitter of gravitational waves.  That is, for a biaxial system where the principal eigenvector of the star's moment of inertia tensor is orthogonal to the stellar rotation axis, implying gravitational wave emission is maximal.  Below we discuss the two most plausible physical scenarios for generating gravitational waves---magnetic field-induced deformations and bar-mode instabilities---and relate these to the ellipticity constraints derived from equation (\ref{internaleps}).


\subsection{Magnetic field-induced deformations}
\label{sec:epsb}
The nascent neutron star born from the binary merger is differentially rotating and therefore likely winds up a strong toroidal component 
of the magnetic field.  Indeed, this expectation is supported by recent numerical simulations \citep[e.g.,][]{rezzolla11,kiuchi14,giacomazzo15}.

Such a strong magnetic field naturally deforms the star, inducing an ellipticity which, for simple stellar models, is well-approximated as~\citep{cutler02},
\begin{equation}
	\epsilon_B\approx 10^{-6}\left(\frac{\left<B_{\rm t}\right>}{10^{15}\,{\rm G}}\right)^2,\label{epsB}
\end{equation}
where $\left<B_{\rm t}\right>$ is the volume-averaged toroidal field strength inside the star \citep[see][for a recent review]{lasky15c}.  

To what extent the magnetically-deformed body can also become an efficient source of gravitational radiation largely depends on geometry:
the magnetic symmetry axis needs to be sufficiently misaligned with respect to the spin axis. Given the mechanism for generating a strong toroidal field, 
the two axes are likely to be almost parallel initially and some additional mechanism is required for making the system more `orthogonal'.
This could be indeed achieved by the so-called spin-flip instability.


\subsection{The spin-flip instability}
\label{sec:epssf}
The symmetry axis of a wound-up toroidal field is expected to be aligned to the rotation axis of the star, implying the induced stellar deformation 
is also rotationally aligned\footnote{An $\alpha$--$\Omega$ dynamo is also likely to operate and further amplify the field \citep[e.g.,][]{duncan92}.  
Large-scale simulations of such evolutions do not exist, and it is unclear what topology the final magnetic field will have.  One can easily imagine such 
a process perturbing the magnetic axis by 10 or 20 degrees, but such a change does not effect significantly the outcome of the calculation in this section.}, 
and the system does not emit gravitational radiation.  However, the energy of such a prolate spheroid is minimised when the magnetic axis is orthogonal 
to the rotation axis. The existence of some dissipative mechanism drives the star to become an orthogonal rotator \citep{mestel72,jones76,cutler02}, and 
therefore an optimal emitter of gravitational waves; this precessional instability is sometimes dubbed a `spin flip'.

The relevant dissipative mechanism in a hot, post-merger remnant is bulk viscosity associated with beta-equilibrium chemical reactions,
in combination with a compressible fluid flow.  The calculation of the spin-flip timescale, $\tau_{\rm sf}$, can be found in Appendix~\ref{app:spinflip}. 
The (approximate) result is:
\be
\tau_{\rm sf} \sim \frac{\rho R^2}{5 | \epsilon_B| \epsilon_\Omega^2 \zeta},
\label{tau_sf0}
\ee
where $\zeta$ is the bulk viscosity coefficient and $\epsilon_\Omega$ is the rotational ellipticity. 

However, when the temperature falls below a critical value, $T_{\rm ad}$, the reaction frequency becomes lower than a fluid element's 
oscillation frequency, and the fluid flow becomes adiabatic and incompressible, and therefore insensitive to bulk viscosity.  
In Appendix~\ref{app:flow} we obtain:
\be
T_{\rm ad}  \approx 9 \times 10^9\,\left(\frac{\rho}{10^{15}\,{\rm g\,cm}^{-3}}\right)^{1/9}\left(\frac{P}{1\,{\rm ms}}\right)^{-1/6} \left ( \frac{\epsilon_B}{10^{-5}} \right )^{1/6}\, ~\mbox{K}.
\label{Tad0}
\ee
Below this temperature (which also marks the maximum value of $\zeta$ and minimum value of $\tau_{\rm sf}$), dissipation in the star is negligible, 
the spin-flip cannot occur, and the star cannot become an orthogonal rotator.  

Simulations of merging neutron stars show that shocks and shear layers drive the maximum temperature of the remnant at the shock interface 
to $T\gtrsim3\times10^{11}\,{\rm K}$, with almost the entire remnant at $T\gtrsim10^{11}\,{\rm K}$ \citep[e.g.,][]{sekiguchi11a,sekiguchi11b,foucart15}.
At this temperature the star cools efficiently by neutrino emission via the modified URCA reactions. The associated cooling profile is \citep[e.g.,][]{shapiro83}
\begin{equation}
	\frac{T(t)}{10^{9}\,\rm K}=\left[\frac{t}{\tau_c}+\left(\frac{10^{9}\,\rm K}{T_{0}}\right)^6\right]^{1/6},
	\label{eqn:mURCA}
\end{equation}
where $T_0=T(t=0)$ and $\tau_c=1.6\times10^{7}\,{\rm s}$.

\begin{figure}
\includegraphics[width=1.0\columnwidth]{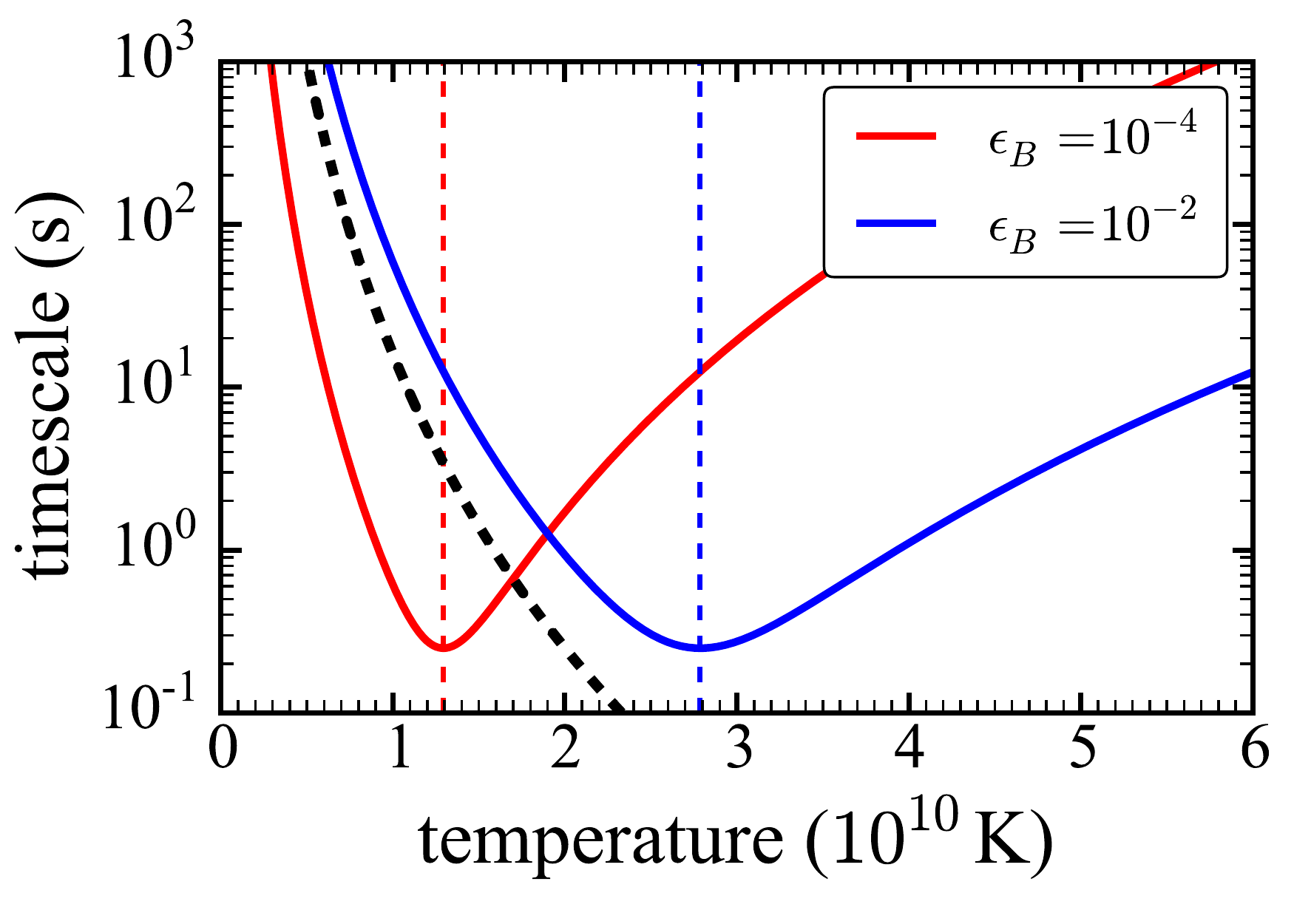}
	\caption{\label{fig:temp} Characteristic timescales for post-merger neutron stars.  The black curve is the mURCA cooling timescale given 
	by equation (\ref{eqn:mURCA}) with $T_0=5\times10^{10}\,{\rm K}$.  The solid curves are the spin-flip timescales, $\tau_{\rm sf}$, 
	for different values of the magnetic field-induced ellipticity, $\epsilon_B$ -- see equation (\ref{tau_sf0}) -- for a star with $P=1\,{\rm ms}$, $R=20\,{\rm km}$ 
	and $\rho=10^{15}\,{\rm g\,cm}^{-3}$. The dashed vertical lines are the respective values of the temperature threshold $T_{\rm ad}$ (eqn.~(\ref{Tad0})) 
	 below which perturbations of the star become adiabatic, implying the spin flip can no longer be driven by bulk viscosity.}
\end{figure}

We plot this as the dashed, black curve in Figure~\ref{fig:temp}, where we have assumed a fiducial $T_0=10^{11}\,{\rm K}$ (note that the temperature
evolution for a system with $T_0\gtrsim 3\times10^{10}\,{\rm K}$ is indistinguishable from the black curve in Figure~\ref{fig:temp} due to 
the $T\propto t^{1/6}$ dependence). In Figure~\ref{fig:temp}, we also show the spin-flip timescale $\tau_{\rm sf}$, for two different values of the magnetic 
ellipticity, $\epsilon_B$. The vertical dashed lines in Figure~\ref{fig:temp} are the values of $T_{\rm ad}$ for the different values of $\epsilon_B$; these 
have been calculated using equation~(\ref{Tad0}). We have assumed a fiducial neutron star with spin period, $P=1\,{\rm ms}$, radius, $R=20\,{\rm km}$ 
and density, $\rho=10^{15}\,{\rm g\,cm}^{-3}$.

Figure~\ref{fig:temp} shows that the cooling curve intersects the spin-flip curve at $T>T_{\rm ad}$ for the $\epsilon_B=10^{-4}$ curve, implying that bulk viscosity 
has enough time to orthogonalise the star before it cools below the temperature $T_{\rm ad}$.  On the other hand, the curve representing $\epsilon_B=10^{-2}$ 
has the cooling timescale intersecting the spin-flip timescale at $T<T_{\rm ad}$, implying orthogonalization cannot occur.  Therefore, if a star is born with too 
large an elliptical deformation, i.e., $\epsilon>\epsilon_{\rm sf}$, orthogonalization cannot occur, and the star will not become an optimal emitter of gravitational waves.  
We calculate this critical ellipticity in Appendix~\ref{app:fail}, 
\be
\epsilon_{\rm sf}\approx5\times10^{-3}\left(\frac{\rho}{10^{15}\,{\rm g\,cm}^{-3}}\right)\left(\frac{P}{1\,{\rm ms}}\right)^{-2}\left(\frac{R}{10\,{\rm km}}\right)^{-2}.
\label{epssf}
\ee

Finally, it is worth noting that the spin-flip may also be rendered inactive due to suppression of bulk viscosity at very high temperatures.  At $T\gtrsim3\times10^{10}\,{\rm K}$, the neutron-star matter is opaque to neutrinos, implying a significantly lower bulk viscosity damping rate (see Appendix~\ref{app:fail} for details).  At such temperatures, the spin-flip timescale becomes longer than the timescales of interest here.  Figure \ref{fig:temp} implies that this effect is only relevant for highly-deformed stars, and therefore does not alter our conclusion that the spin-flip will only occur for stars with ellipticities below the critical ellipticity given in equation (\ref{epssf}).  We note that these calculations should be backed up by more robust analyses, rather than the back-of-the-envelope analysis undertaken herein, however this is beyond the scope of the present paper.


\subsection{Bar-mode deformations}
\label{sec:bar}
While magnetic field-induced ellipticities rely on the orthogonalisation of the stellar spin and magnetic axes to become potent
emitters of gravitational radiation, it is not the only gravitational wave emission mechanism that may operate in a post-merger neutron star.  
Another viable possibility is that of unstable stellar pulsations and in particular the unstable $f$-mode, otherwise known as the bar-mode instability \citep{andersson03,corsi09a}.  These come in two flavours: dynamical and secular instability; the latter is the celebrated 
Chandrasekhar-Friedman-Schutz (CFS) gravitational wave-driven instability \citep{chandrasekhar70,friedman78}.

The onset and growth timescale of the $f$-mode instability is determined by the system's $\beta \equiv T/|W|$ ratio (kinetic to gravitational binding
energy) with the dynamical instability requiring a higher $\beta$ value than the secular one to set in. In realistic neutron stars, the threshold 
$\beta_{\rm dyn} \approx 0.27$ for the onset of the dynamical instability can be reached only in conjunction with the presence of differential rotation. 
Hence, this instability is associated with the (very) early phase of the post-merger remnant's life and should be suppressed once differential rotation 
is quenched by the magnetic field---the relevant timescale for this to happen is the Alfv\'en timescale \citep{shapiro00} which, for these systems, is $\ll1\,{\rm s}$. 
 
More relevant to our discussion is the secular $f$-mode instability: in this case the threshold $\beta_{\rm sec} \approx 0.14$ can be exceeded
even in rigidly rotating systems and therefore the instability could be active on a spin-down timescale.  In fact, a massive and rapidly spinning remnant
 may ``optimise" the $f$-mode instability, in the sense that the growth timescale could be unusually short and that
 the dominant multipole is the quadrupolar $\ell=m=2$.  This is suggested by the recent work of \citet{doneva15} who have reported a timescale 
 $t_{\rm grow} \sim 10-100\,\mbox{s}$; this is much shorter than previous results in the literature \citep{gaertig11,doneva13,passamonti13} 
 and could clearly be of relevance for the dynamics of the system during the afterglow X-ray plateau phase. The main prerequisite  
 is that the system has cooled below a  temperature $\sim 10^{10}\,\mbox{K}$ so that the instability is not suppressed by bulk viscosity. 
 This should indeed happen a few seconds after the merger, see the cooling curve in Fig.~\ref{fig:temp}.

In order to fit the $f$-mode instability to our short GRB model we need to estimate the stellar ellipticity induced by the unstable mode. 
This can be achieved with the help of the following argument. 

The mode energy can be approximated by its kinetic part, i.e. $E_{\rm mode} \approx (1/2) \rho \omega^2 \xi^2 V$ where $V$ is the stellar 
volume and $\xi$ is the average mode amplitude. For the change in the stellar radius we have $\delta R \sim \xi$ and for the mode frequency we can use 
the result for a non-rotating star, $\omega^2 \approx 2GM/R^3$. The resulting ellipticity is:
\be
\epsilon_f \approx \frac{2\delta R}{R} \sim  \left ( \frac{E_{\rm mode}}{M c^2} \right )^{1/2} \left ( \frac{c^2R}{GM} \right )^{1/2}.
\label{eccf}
\ee
A more rigorous calculation, based on the $f$-mode eigenfunction and frequency of a non-rotating uniform density Newtonian star,
returns a result within a factor of order unity from our back-of-the-envelope formula (\ref{eccf}).

Recent work on the non-linear dynamics of unstable $f$-modes suggests that the saturation energy 
can reach a maximum value $E_{\rm mode} \approx 10^{-6} M c^2$ although a typical value of this quantity could be
10-100 times lower~\citep{pnigouras15,doneva15}. Thus the uncertainty attached to $E_{\rm mode}$  could easily outweigh the error due 
to the approximate nature of (\ref{eccf}). 

For canonical neutron star parameters our result (\ref{eccf}) predicts a maximum $f$-mode-induced ellipticity $\epsilon_f \sim 10^{-3}$. 
As clearly seen in Fig.~(\ref{ellipticity}), this could be comparable to the observational upper limits for the ellipticity and therefore 
gravitational wave-unstable $f$-modes could, in principle, drive the spin down of the post-merger neutron star. 


\subsection{Inertial modes}
\label{sec:inertial}

The second class of oscillation modes that could undergo the CFS gravitational wave-driven instability is that of inertial $r$-modes 
\citep[see][for a review]{andersson01}.  Unlike the previous case of the bar $f$-mode, the quadrupolar $r$-mode
(which is the most unstable one) cannot be represented as an induced ellipticity; nevertheless we can still study the spin evolution of
the system under $r$-mode radiation. 

After the mode has saturated at its maximum amplitude, $\alpha_{\rm max} \ll 1$, the stellar spin evolution is governed by \citep{owen98}
\be
\dot{\Omega} \approx -2Q \alpha^2_{\rm max}  \frac{\Omega}{\tau_{\rm gw}},
\ee
where $Q$ is a stellar structure-dependent parameter (see below) and $\tau_{\rm gw}$ is the mode's growth timescale. Since $\tau_{\rm gw} (t) \sim \Omega(t)^{-6}$,
it is easy to see that the $r$-mode-dominated spin-down leads to a luminosity profile,
\be
L (t) = L_{\rm em,0}  \left ( 1 + \frac{t}{t_{\rm gw}} \right )^{-2/3}.
\ee
The characteristic spindown timescale $t_{\rm gw}$ is found to be:
\begin{align}
t_{\rm gw} = &\frac{ \tau_{\rm gw} (0)}{12 Q \alpha^2_{\rm max}}\notag\\ 
\approx &5\times 10^9\, \left(\frac{M}{1.4\mbox{M}_\odot}\right)^{-1}\left(\frac{R}{10\,\mbox{km}}\right)^{-4} \left ( \frac{P}{1\,\mbox{ms}} \right) ^6 \left ( \frac{10^{-4}}{\alpha_{\rm max}} \right )^2\, \mbox{s},
\end{align}
where we have used $Q \approx 0.092$ calculated for an $n=1$ Newtonian polytrope and have normalised $\alpha_{\rm max}$ 
according to the predictions of nonlinear calculations \citep{bondarescu07, bondarescu09}. 

The preceding results suggest that the $r$-mode instability is not likely to play a significant role in the 
evolution of the post-merger remnant. This is because (i) the predicted X-ray late time tail is much shallower than what the data
suggest and (ii) the spin down timescale is significantly longer than the observed plateaus. 
In addition to these, it is worth noting a theoretical complication: the dynamics of unstable $r$-modes in strongly magnetised neutron
stars remains poorly understood, perhaps involving the winding up of the stellar magnetic field and the saturation of the 
instability itself \citep[e.g.,][]{rezzolla00,rezzolla01a,rezzolla01b}.


\section{Observational Constraints}
\label{ellipticity_constraints}
\citet{rowlinson13} analysed the gamma- and $X$-ray light curves of 43 short GRBs observed with the {\it Swift} satellite.  
Of this sample, only eight have confirmed redshifts and are well fit by the magnetar model.  Our analysis relies on having confirmed redshifts 
to give reliable intrinsic luminosities so that we can make inferences about the nascent neutron star.  This sample of eight GRBs are the same as those 
used in \citet{lasky14} to determine equation of state properties of the newly born neutron stars.

\begin{figure}
\includegraphics[width=1.0\columnwidth]{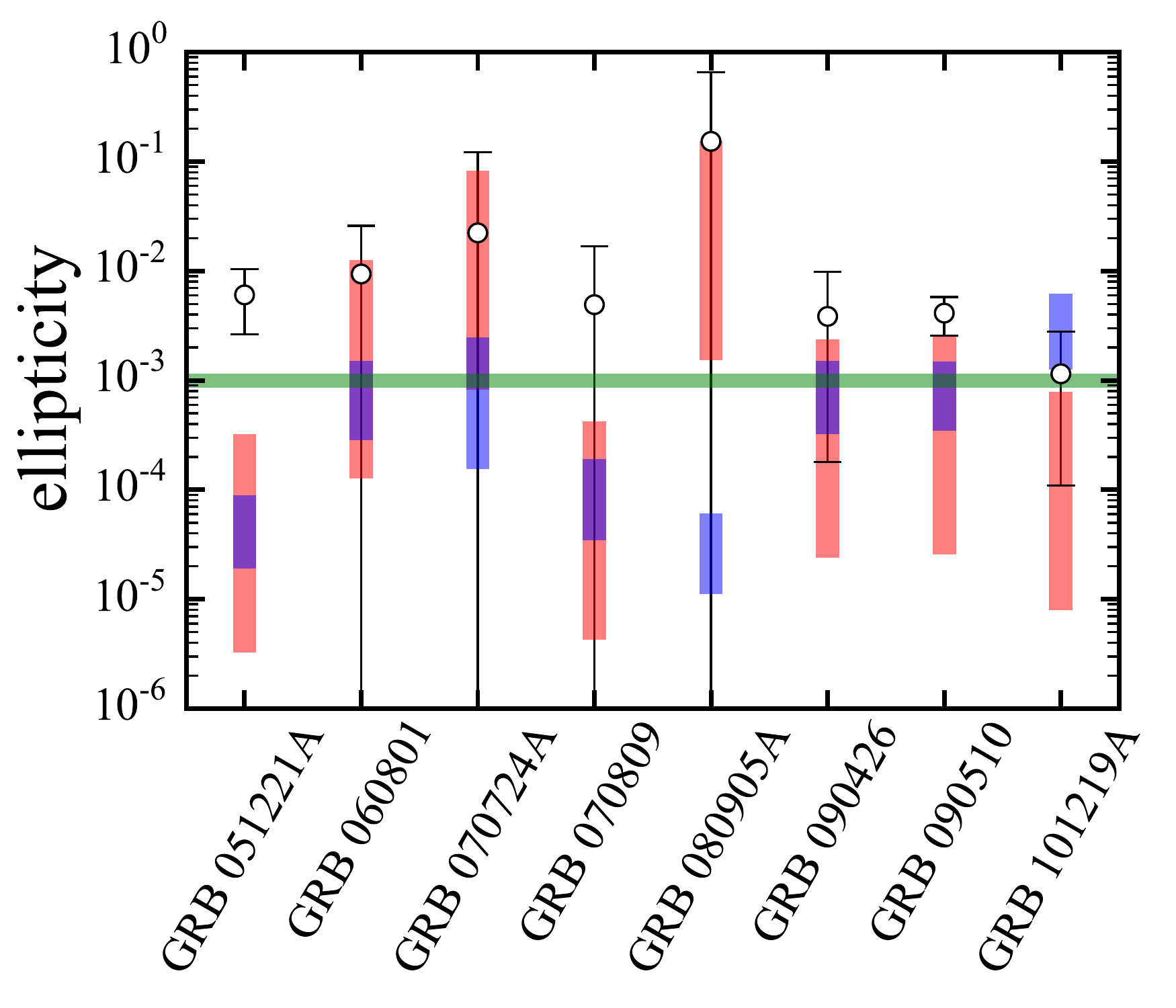}
	\caption{\label{ellipticity}Upper limit on the ellipticity for the eight short GRBs in our sample.  The black points with error bars are the observationally inferred upper limits on the ellipticity, $\epsilon_{\rm obs}$, constrained by equation~(\ref{internaleps}) assuming an efficiency $\eta=0.1$.  The red bars show the range of expected magnetic field-induced ellipticities, $\epsilon_{B}$, given  by equation (\ref{epsB}), the blue bars show the range of maximal ellipticities for which the spin flip can take place, $\epsilon_{\rm sf}$ from equation (\ref{epssf}) (i.e., a spin flip can only occur if the ellipticity is {\it smaller} than the ellipticity given in blue), and the horizontal green line is the approximate bar-mode saturation ellipticity given in Section \ref{sec:bar}.}  \end{figure}

In Figure~\ref{ellipticity} we plot all of the derived ellipticity constraints from Section \ref{sec:epsall} for each of the eight short GRBs in our sample.  
In black, we plot the observational upper limits on the ellipticity, $\epsilon_{\rm obs}$, from equation (\ref{internaleps}) assuming an efficiency of $\eta=0.1$, 
where the error bars take into account the uncertainty in the measurements of the plateau luminosity and timescale.  In red we plot the range of possible 
magnetic field-induced ellipticities, $\epsilon_B$, from equation (\ref{epsB}) assuming $\left<B_t\right>=\alpha B_p$, where $B_p$ is the observationally 
inferred poloidal magnetic field strength for each GRB, and $1\le\alpha\le10$; calculations of neutron star hydromagnetic equilibria suggest this is a plausible 
range for $\alpha$ \citep[e.g., ][]{braithwaite09,akgun13}.  In blue we plot the critical ellipticity above which the spin flip instability cannot take place, 
$\epsilon_{\rm sf}$, from equation (\ref{epssf}), where the lower limit assumes the lower bound on the protomagnetar's spin period and $R=10\,{\rm km}$, 
and the upper limit assumes the upper bound on the protomagnetar's spin period and $R=20\,{\rm km}$.  That is, the range showed as the blue region 
accounts for both the uncertainty in measurement of the neutron stars spin, as well as uncertainties associated with the neutron star equation of state.  
Finally, the horizontal green line in Figure~\ref{ellipticity} is the estimated maximum $f$-mode ellipticity, $\epsilon_{\rm f}$, from Section \ref{sec:bar}.

Most GRBs in Figure~\ref{ellipticity} show the observationally-derived upper limit on the ellipticity (i.e., the black points) as larger than the magnetic 
field-induced deformations (red bars), implying the {\it actual} ellipticity of the body is likely to be smaller than the observationally-inferred upper limit.  
For example, observations of GRB 051221A indicate the ellipticity of the star can be as high as $\epsilon\lesssim10^{-2}$ without affecting the shape of the light curve.  
However, the inferred poloidal component of the magnetic field indicates that $3\times10^{-6}\lesssim\epsilon\lesssim3\times10^{-4}$.  For this GRB, the error bars 
associated with the observationally-inferred and magnetic field-induced ellipticities are distinct; a majority of other GRBs in our sample show significant overlap 
between these two quantities, implying it is possible that the observationally-derived upper limit could be realised by magnetic field deformations alone.  
In these cases, the important question becomes: can these systems become optimal emitters of gravitational waves from magnetic-field induced deformations?

There are a number of scenarios that are plausible given the four different values of ellipticity shown for each GRB in Figure~\ref{ellipticity}.  We break this into two categories; firstly based on only magnetic-field induced ellipticities, and secondly on the $f$-mode instability.  There is a mix of systems in our sample where one would expect the bar-mode to dominate gravitational wave emission (GRBs 0512121A, 070809, 101219A), and where the gravitational wave emission can be dominated by magnetic-field induced instabilities.  In the former case, one expects the maximum ellipticity to be given by ${\rm min}\{\epsilon_{f},\,\epsilon_{\rm obs}\}$, while the latter would expect ${\rm min}\{\epsilon_{\rm B},\,\epsilon_{\rm obs}\}$ Details of this inference are outlined below.


\subsection{Magnetic-field induced ellipticity}
We summarise the various situations as follows:
\begin{itemize}
	\item $\epsilon_{\rm obs}>\epsilon_{B}>\epsilon_{\rm sf}$:  As $\epsilon_B>\epsilon_{\rm sf}$, the magnetic-field induced ellipticity is too large, and 
	therefore necessarily prohibits the spin-flip from taking place.  Only for GRB 080905A is $\epsilon_B$ necessarily larger than $\epsilon_{\rm sf}$ --- in this case, 
	despite the observationally-derived ellipticity upper limit being large, $\epsilon_{\rm obs}\sim0.1$, this system cannot optimally emit gravitational waves as the spin-flip 
	instability cannot take place.  It is possible that this condition also holds for many of the other GRBs, although the error bars are too large to determine this uniquely 
	for any other GRB in our sample.
	\item $\epsilon_{\rm obs}>\epsilon_{\rm sf}>\epsilon_{\rm B}$:  If the magnetic-field induced ellipticity is {\it less} than the critical spin-flip instability, then the spin-flip 
	can take place, the system can become an orthogonal rotator on a relatively short timescale, and gravitational waves can be optimally emitted.  In this case, the 
	gravitational wave luminosity is dictated by $\epsilon_B$.
	\item $\epsilon_{\rm sf}>\epsilon_{\rm obs}>\epsilon_{\rm B}$:  In this case, the spin-flip instability can always take place, and the gravitational wave emission is 
	likely given by the value of $\epsilon_B$.
	\item $\epsilon_{\rm sf}>\epsilon_{\rm B}>\epsilon_{\rm obs}$:  We suggest that this case is unphysical.  Although in principle the spin-flip instability can take place 
	(i.e., $\epsilon_{\rm sf}>\epsilon_{B}$), if the system emitted gravitational waves as an orthogonal rotator with  $\epsilon_B$, this would necessarily have 
	induced a noticeable change in the X-ray light curve. 
\end{itemize}


\subsection{Bar-mode instability}

The observationally-inferred ellipticity upper limits shown in Fig.~\ref{ellipticity} can also be viewed as upper limits on
the ellipticity $\epsilon_f$ of the secular bar-mode and therefore on the instability's saturation energy [see Eqn.~(\ref{eccf})],
provided of course that the $f$-mode instability does operate in these systems. 

The horizontal green line in  Fig.~\ref{ellipticity} represents the bar-mode ellipticity for the theoretically predicted maximum saturation energy 
$E_{\rm mode} \sim 10^{-6} M c^2$ (see Section \ref{sec:bar}). In all cases, this falls below the observational upper limits (although error bars on this quantity often 
fall below $\epsilon_{f}$), implying an overall consistency between short GRB observations and theoretical bar-mode calculations.

Similar to what we have seen for magnetic field-induced ellipticities, the case $\epsilon_{\rm obs}<\epsilon_{f}$ is deemed unphysical
as the $X$-ray light curve would take on a markedly different evolution. For this to happen, $\epsilon_f$ would have to increase to a value
$\epsilon_f \sim 10^{-2} $ which translates to a mode energy as high as $E_{\rm mode} \sim 10^{-4} Mc^2$.


\section{Gravitational Wave Constraints}\label{sec:GWcons}
In the previous section we spent considerable effort detailing theoretical scenarios that could effect the overall ellipticity, and hence gravitational wave luminosity that could potentially come from the nascent neutron star.  However, in this section we focus on an {\it absolute} gravitational wave upper limit from these neutron stars, where we remain completely agnostic about the theoretical framework underpinning the gravitational wave emission.  That is, we only look at the empirically-derived gravitational wave upper limit assuming that the maximum ellipticity the star can possibly has is $\epsilon_{\rm obs}$ from equation (\ref{internaleps}).  This is the most conservative gravitational wave upper limit we can derive given that, were the gravitational wave luminosity to be larger than this value, it would effect the $X$-ray light curve following the prompt emission.  

The gravitational wave strain for a rotating neutron star at distance, $d$ is,
\begin{align}
	h(t)=\frac{4G}{c^4}\frac{\epsilon I\Omega(t)^2}{d}.
\end{align}
The optimal matched filter signal-to-noise ratio is given by
\begin{align}
	\rho_{\rm opt}^2=\int_{f_i}^{f_f}df\frac{\tilde{h}(f)^2}{S_h(f)},\label{SNR}
\end{align}
where $\tilde{h}(f)$ is the Fourier transform of $h(t)$, $S_h(f)$ is the noise power spectral density of the detector, and $f_i$ and $f_f$ are respectively the initial and final gravitational wave frequencies.  The stationary phase approximation implies $\tilde{h}(f)^2=h(t)^2\left|dt/df\right|$, where the gravitational wave frequency evolution, $df/dt$, is derived directly from equation (\ref{spindown}).

Equation (\ref{SNR}) assumes an optimal matched filter, which is most likely not feasible for the detection of such long-lived (i.e., $\gtrsim 10\,{\rm s}$) transient signals.  In reality, alternative algorithms are used that require the strain be increased by a factor of a few (equivalently, the distance must be reduced by a factor of a few) for the pipelines to make a detection with equivalent false alarm and false dismissal rates as seen with the optimal matched filter \citep[e.g.,][]{thrane13a,thrane15a,coyne15}.  In the spirit of deriving strict upper limits on the signal-to-noise ratio, we persist with equation (\ref{SNR}) for the remainder of this article, but note that realistic detection algorithms will reduce the signal-to-noise ratio by some non-negligible factor.

In the limit where gravitational wave emission dominates spindown, $\tilde{h}(f)$ can be expressed as 
\begin{align}
\tilde{h}(f)=&\frac{1}{d}\sqrt{\frac{5GI}{2c^3f}}\\
	\approx&2.6\times10^{-25}\left(\frac{I}{10^{45}\,\mbox{g\,cm}^2}\right)^{1/2}\,\left(\frac{d}{100\,{\rm Mpc}}\right)^{-1}\left(\frac{f}{1\,{\rm kHz}}\right)^{-1/2}\,{\rm s}.\label{eq:htilde}
\end{align}
Equation (\ref{eq:htilde}) is independent of the neutron star ellipticity, but only depends on the assumption that the stellar angular frequency evolves according to equation (\ref{spindown}) with negligible dipole radiation \citep[e.g.,][]{owen98,dallosso15}.  An upper limit on the signal-to-noise is therefore controlled by the distance to the source, as well as the initial and final frequencies of the system.  

\begin{figure}
\includegraphics[width=1.0\columnwidth]{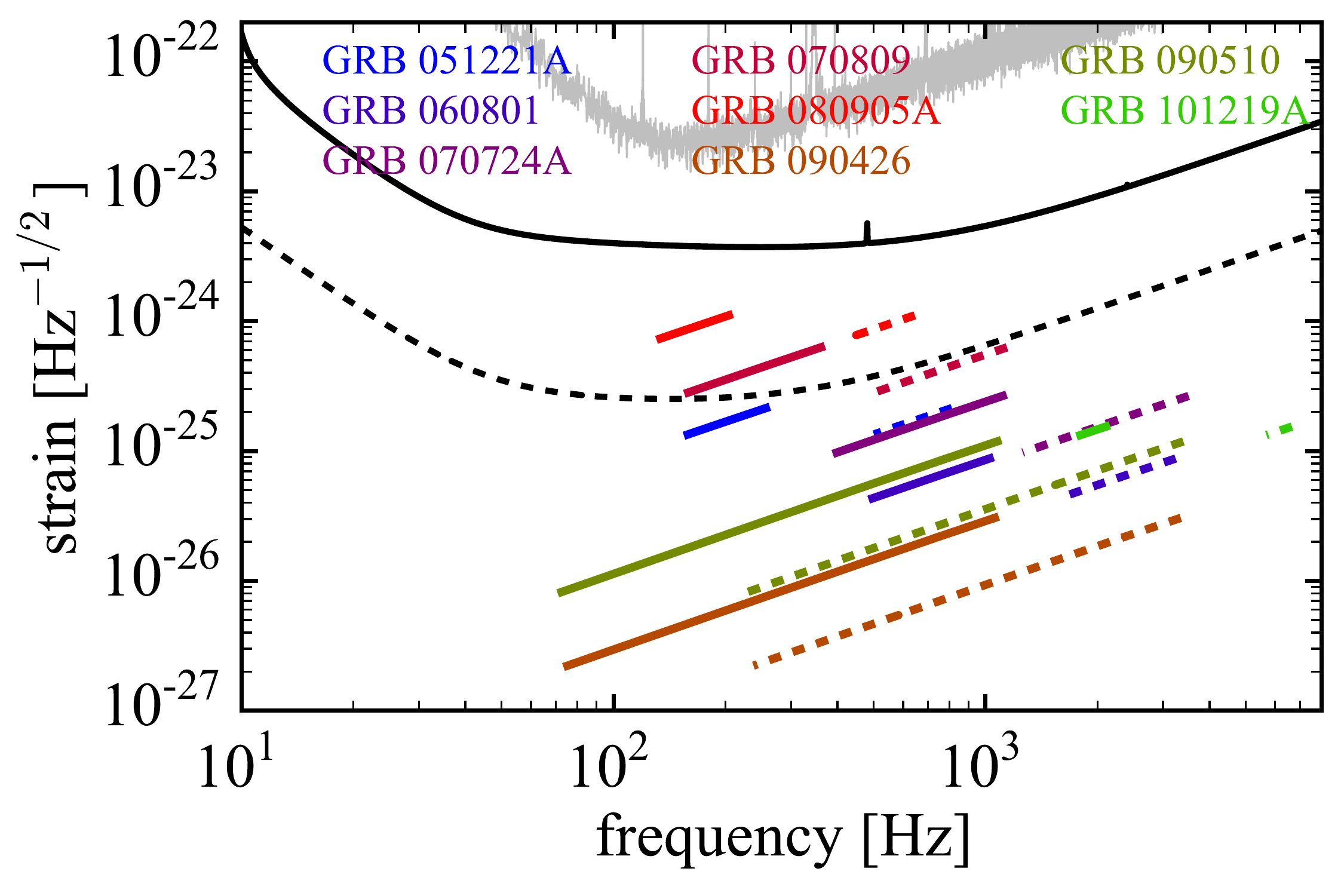}
	\caption{\label{strain_curve}Gravitational wave strain evolution for the eight short GRBs in our sample assuming an efficiency of $\eta=1$ (solid, coloured curves) and $\eta=0.1$ (dashed, coloured curves).  For those GRBs that are not observed to collapse, we illustrate the gravitational wave signal lasting $10^5$ seconds -- see text.  The solid grey, solid black and dashed black curves are the noise power spectral densities, $S_h(f)$, for the S5 run of initial LIGO, projected sensitivity for aLIGO and the Einstein Telescope respectively.}
\end{figure}

In Figure~\ref{strain_curve} we plot the gravitational wave strain, $\tilde{h}(f)$, for the eight short GRBs in our sample as well as the noise power spectral density, $S_h(f)$, for the S5 run of initial LIGO \citep[][grey curve]{PSD:iLIGO} and the projected sensitivities for aLIGO at design sensitivity \citep[][solid, black curve]{PSD:aLIGO} and projected sensitivity for the Einstein Telescope \citep[][dashed, black curve]{hild11}.  For each GRB, we plot the gravitational wave strain evolution assuming $\eta=1$ (solid, coloured curves) and $\eta=0.1$ (dashed, coloured curves).  For those GRB remnants that are observed to have a sharp decay in the light curve, we assume gravitational wave emission lasts from the time of the prompt emission to the time of the sharp decay, interpreting this moment as the collapse of the supramassive neutron star to a black hole.  For the other short GRBs that do not show sharp decays in their light curves, we assume for illustrative purposes that the gravitational wave signal lasts $10^{5}$ seconds.  Although four of the GRBs shown in Figure~\ref{strain_curve} have the same observation time, the length of the curves in these figures also depends on the initial spin frequency through equation (\ref{eq:htilde}).

\begin{figure}
\includegraphics[width=1.0\columnwidth]{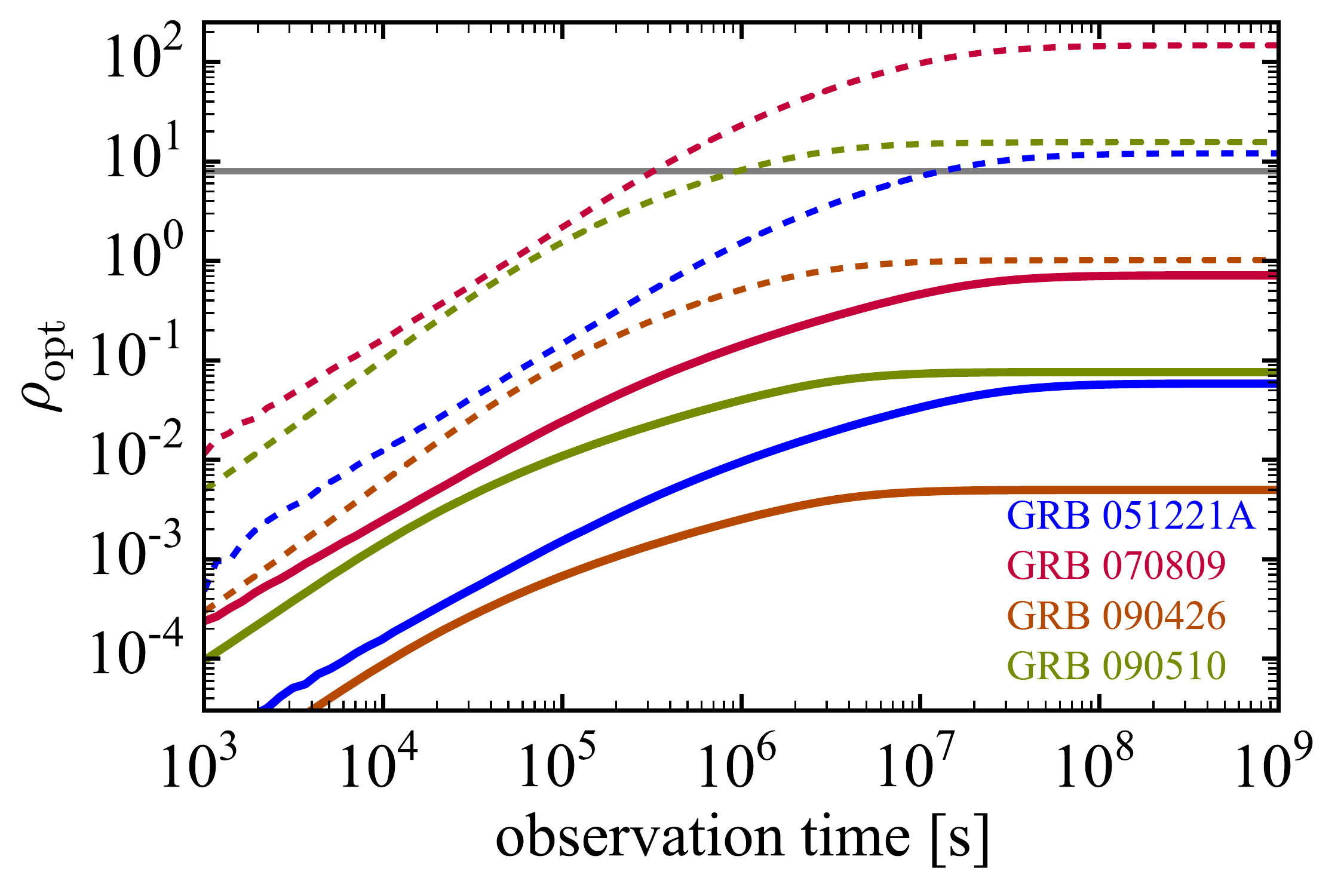}
	\caption{\label{fig:SNR}Optimal matched filter signal-to-noise ratio, $\rho_{\rm opt}$ from equation (\ref{SNR}) as a function of observation time for Advanced LIGO (solid curves) and the Einstein Telescope (dashed curves).  The horizontal grey line shows a nominal value of $\rho_{\rm opt}$ that traditionally represents a detection---although note the curves shown are absolute upper limits, and should not be considered realistic.}
\end{figure}

In Figure~\ref{fig:SNR} we plot the optimal, matched filter signal-to-noise ratio, $\rho_{\rm opt}$, as a function of observing time for both aLIGO at design sensitivity (solid, coloured curves) and the hypothetical Einstein Telescope (dashed, coloured curves).  We only plot the four GRBs from the sample that do not have a steep decay phase in their X-ray light curve.  That is, in the magnetar model, these GRBs gave birth to stable neutron star remnants.  The curves plateau after some observation time primarily because the gravitational wave frequency has moved out of the sensitivity band of the detectors.  This plot paints a bleak picture in terms of detection for aLIGO but, at first blush, shows some promise for third-generation detectors such as the Einstein Telescope.  However, we remind the reader that the curves in Figure~\ref{fig:SNR} not only represent the most optimistic upper limit in terms of gravitational wave energy loss from the nascent neutron star, but are also incredibly optimistic in terms of the data analysis algorithms for making the detection.  In particular, the curves in Figure~\ref{fig:SNR} are calculated assuming an optimal matched filter, which is not computationally feasible for signals lasting $\sim 10^{5}$ s.  As mentioned, realistic detection strategies will reduce $\rho$ by some non-negligible factor \citep[e.g.,][]{thrane13a,thrane15a}.

\section{Conclusion}
\label{sec:conclusions}
There is increasing observational and theoretical support for the millisecond magnetar model for short GRBs, in which the merger of two neutron 
stars gives birth to a new, rapidly rotating, highly magnetised neutron star.  Such exotic conditions are conducive to large deformations that will cause 
the neutron star to emit a large amount of gravitational radiation.  In this paper, we use the shape of the observed X-ray light curves following short GRBs 
to place upper limits on the gravitational wave emission from these stars, showing that they are unlikely detectable in second-generation ground-based 
gravitational wave interferometers such as aLIGO.  Moreover, existing observations suggest these systems are only marginally detectable by third-generation 
gravitational wave interferometers. Only short GRBs occurring much closer than the typical distances associated with such events 
 may stand a good chance of gravitational wave detection 

Our gravitational wave detection prospects are prima facie in conflict with the predictions of \citet{dallosso15} and \citet{doneva15}.  However, those authors calculate gravitational wave signal-to-noise ratios assuming a fiducial distance to the merger of $d=33.5$ and 20 Mpc respectively.  The nearest GRB in the entire \citet{rowlinson13} sample lies at luminosity distance $d\approx500$ Mpc (GRB 061201, with redshift $z=0.111$), while the closest in our sub-sample is GRB 080905A, with $d\approx570$ Mpc.  Signal-to-noise ratio scales as $1/d$ [equations~(\ref{SNR}) and (\ref{eq:htilde})], implying an otherwise equivalent GRB an order of magnitude closer has a factor of ten larger $\rho$.  Even this extra factor of ten in $\rho$ renders our predictions pessimistic for aLIGO (see Figure~\ref{fig:SNR}).  This is especially true considering $\rho_{\rm opt}$ assumes an unrealistic, optimal matched filter detection statistic---see discussion below Equation~(\ref{SNR}).

Despite the pessimism of our gravitational wave upper limits, one should be very optimistic about the prospects of learning physics from these 
observations in general.  Such observations already suggest, assuming the standard post-merger magnetar model, that the equation of state of neutron star matter 
needs to be adjusted to include non-rotating maximum masses as high as $\sim2.3\,M_\odot$ \citep{fan13, lasky14, lu15}.  Moreover, in this paper we 
have used the observations to constrain both the ellipticity and $f$-mode saturation amplitude~\citep[see also][]{gao15}.  
For example, we have shown that magnetic field-induced deformations of the nascent neutron star do not generally effect the spindown of the neutron 
star as they are either too insignificant, or the spin-flip instability is too slow to operate.  Also, while the bar-mode instability {\it can} operate, it too does not 
sufficiently deform the star to impinge significantly on the spindown.  

An exciting future prospect is that of parameter estimation associated with future gravitational wave observations of the inspiral phase of the compact objects.  
One of the holy grails of gravitational wave astronomy is low-latency gravitational wave and electromagnetic observations of binary neutron star inspiral and 
merger phases \citep[e.g., see][]{chu15}.  Such simultaneous observations would allow unprecedented parameter estimation in terms of the GRB 
progenitor \citep{aasi13}, which could then be used to inform the modelling and understanding of the post-merger remnant.  Such a programme has already 
been shown to enhance our understanding of the neutron star equation of state \citep{lasky14}, but will undoubtably lead to other advances.  
For example, such multimessenger observations, coupled with hydrodynamic simulations, could provide an accurate measurement of $T/|W|$, and hence 
a deeper understanding of instabilities in the newly born neutron star --- such studies are currently underway.

The millisecond magnetar model employed in this work may not be the final word on the subject. Recent work~\citep{rezzolla15,ciolfi15} 
has highlighted the importance of accounting for the presence of the ejecta produced during the merger. The outflowing matter could effectively 
trap light temporarily and time-reverse the sequence of the main gamma flash (produced during the collapse of the remnant to a black hole) 
and the spin down-powered X-ray signal. Calculating the total gravitational wave emission in these aptly called  `two-wind' or `time reversal'  
scenarios~\citep{rezzolla15,ciolfi15}  is a difficult task  given the uncertainty in the physics underpinning the delay between the merger 
of the binary and the launch of the jet. More specifically, the identification of the plateau timescale with the spin down timescale 
cannot be made anymore, thus removing with one stroke the backbone of our calculation.  
However, it is also worth noting that these models now face a difficult hurdle in explaining the formation of the jet from 
the collapse of the black hole as it has been shown that these produce insufficient debris disks \citep{margalit15}.


\appendix

\section{The `spin-flip' instability}
\label{app:spinflip}

This appendix provides a detailed discussion of the `spin-flip' instability that could take place in systems like magnetically 
deformed rotating neutron stars. Under the action of this instability, which was first discovered by \citet{mestel72} and \citet{jones76}  
in their study of the oblique rotator dynamics, the stellar spin axis tends to become orthogonal to that of the magnetic field on a viscous
timescale. The more modern discussion of the instability in the context of gravitational wave emission 
from rapidly spinning neutron stars was initiated by~\citet{cutler02}.  

It is straightforward to demonstrate the existence of this instability in a rotating and magnetic fluid body. Such a system will
generically find itself in a state of free precession when the spin vector $\mathbf{\Omega}$ is misaligned with respect to
the magnetic field's symmetry axis. If $\chi$ is the angle between these two directions, then the total rotational kinetic energy takes
the form~\citep{cutler02}:
\be
E_{\rm rot} = \frac{J^2}{2 I_0} \left (\, 1-\epsilon_\Omega -\epsilon_B \cos^2 \chi  \, \right ),
\label{KE1}
\ee 
where $J$ is the (conserved) total angular momentum, $I_0$ is the is the moment of inertia of the spherical body
 (i.e. without rotation and magnetic field) and 
 \be
\epsilon_\Omega = \frac{\Delta I_\Omega}{I_0}, \qquad \epsilon_B = \frac{\Delta I_B}{I_0},
\ee
are dimensionless ellipticities produced by the centrifugal and magnetic forces, respectively.

The rotational deformation is always oblate in shape i.e. $\epsilon_\Omega >0$. On the other hand, $\epsilon_B$ 
can be positive (if the magnetic field is predominantly poloidal with respect to its symmetry axis) or negative (if the magnetic 
field is predominantly toroidal). In the latter case the shape of the magnetic deformation is prolate rather than oblate.

It is evident from (\ref{KE1}) that a system with $\epsilon_B > 0$ minimises its energy at $\chi=0$, i.e. when the body is an
aligned rotator. In the opposite case of a prolate body, $\epsilon_B < 0$, the minimum energy state is that of an 
orthogonal rotator, $\chi=\pi/2$, the system thereby acquiring the optimal geometry for gravitational wave emission. 
Both minimum energy configurations are also no-precession states.   

The spin-flip is a secular type of instability, driven by the coupling to some dissipative mechanism.
The exact nature of this mechanism also determines the spin-flip's characteristic timescale which is a key quantity of interest for 
this paper. This is discussed in the remainder of this appendix. 


\subsection{The spin-flip timescale}
\label{app:timescale}

For the hot post-merger remnants considered in this paper the main dissipative mechanism 
that could drive the spin-flip instability is bulk viscosity associated with beta equilibrium chemical 
reactions (the high temperature in such a systems implies a negligible shear viscosity).  It is also worth noting
that energy lost to gravitational waves cannot produce the desired effect as it always tends to make the system an aligned rotator on a very long timescale~\citep{cutler01}.

Viscosity is  effectivelly ``switched off" as long as a precessing fluid body is modelled under the (standard) assumption of
rigid-body rotation as this type of motion does not lead to a viscous force. However, as~\citet{mestel72} have shown, 
this assumption is at odds with the system being in hydrostatic equilibrium. The problem is fixed by means of 
a ``secondary" flow $\delta \mathbf{v}$ (as measured in the body's rotating frame) which is a precession-driven
oscillation whose role is to enforce hydrostatic equilibrium during precession. 
This flow is expected to be much smaller than the main rotational velocity and vanish 
when the system ceases to precess. Indeed, as we discuss below, the secondary 
flow amplitude can be approximated as [see Eqn.~(\ref{dv_ampl})]:
\be
\delta v \sim \sin2\chi \cos\chi\,\epsilon_\Omega |\epsilon_B| \Omega R,
\label{dv1}
\ee
and its frequency is equal to the precession frequency,
\be
\omega_{\rm pr} = |\epsilon_B| \Omega\cos\chi.
\ee
Apart from enforcing hydrostatic equilibrium,  $\delta \mathbf{v}$ also couples precession to viscosity,
thus opening the way for the spin-flip instability. The relevant viscous timescale is given by the 
integral expression~\citep{ipser91}:
\be
\tau_{\rm bv} = \frac{2 E_{\rm pr}}{| \dot{E}_{\rm pr}|} =  2 E_{\rm pr} 
\left (  \int dV \zeta |\nabla \cdot \delta \mathbf{v}|^2 \right )^{-1},
\label{diss1}
\ee
where $\zeta$ is the bulk viscosity coefficient and $E_{\rm pr}$ is the precessional kinetic energy [see Eqn.~(\ref{KE1})],
\be
E_{\rm pr} =  \frac{J^2}{2I_0} |\epsilon_B| \cos^2 \chi. 
\label{Epr}
\ee 
As we show below, the secondary flow is expected to be compressible (i.e. $\nabla \cdot \mathbf{\delta v} \neq 0$)
as long as the temperature exceeds a threshold $T > T_{\rm ad}$ above which the flow is
non-adiabatic (in the sense that the composition of a perturbed fluid element can change during a precession period). 
The system becomes oblivious to bulk viscosity for $ T< T_{\rm ad}$ as a result of the flow being adiabatic and incompressible 
and therefore the spin-flip instability does \textit{not} take place.  This key point, which is central to our discussion here,
has been overlooked in previous work on this subject.

We can combine (\ref{dv1}) and (\ref{diss1}) and obtain an approximate result for the damping timescale
$\tau_{\rm bv}$.  The assumption of a uniform density star, $\rho = 3M/4\pi R^3$, also implies a uniform coefficient $\zeta $ and 
$I_0 \approx (2/5) M R^2$.  After approximating $\nabla \cdot \delta \mathbf{v} \sim \delta v/R$, we  obtain
\be
\tau_{\rm bv} \sim  \frac{2R^2\rho}{5\sin^2 2\chi |\epsilon_B| \epsilon^2_\Omega \zeta}. 
\ee
The timescale $\tau_{\rm sf} $ for the orthogonalisation (or alignment) of the spin and magnetic axes
can be found with the help of
\be
\dot{E}_{\rm pr} = \frac{J^2}{2I_0} \epsilon_B \sin 2\chi \dot{\chi}.
\ee
Then,
\be
\tau_{\rm sf} \equiv \frac{\sin\chi}{|\cos\chi  \dot{\chi}|} \quad \to \quad \tau_{\rm sf} = \tan^2 \chi\,  \tau_{\rm bv}.
\label{tau_chi1}
\ee
Assuming a small initial inclination, $\cos \chi \approx 1$, we finally obtain the spin-flip timescale
\be
\tau_{\rm sf} \sim  \frac{R^2\rho}{10 |\epsilon_B| \epsilon^2_\Omega \zeta}. 
\label{tau_chi2}
\ee
We can calculate this timescale with the help of the approximate numerical estimates for the
rotational and magnetic deformations~\citep{cutler02},
\be
\epsilon_\Omega \approx 0.3 \, P_{\rm ms}^{-2}, \quad 
\epsilon_B \approx   \pm 10^{-6}\, \langle B \rangle_{15}^2,
\ee
where $P_{\rm ms} = P/1\,\mbox{ms}$ is the normalised spin period  and $\langle B \rangle_{15} = \langle B \rangle/10^{15}\, \mbox{G}$ is the 
normalized volume-averaged internal magnetic field (the different signs correspond to a dominantly poloidal/toroidal magnetic field).
The resulting timescale $\tau_{\rm sf}$ as a function of temperature is shown in Figure~\ref{fig:temp}, together with the mURCA cooling timescale.

The bulk viscosity coefficient is, in general, a function of the fluid's oscillation frequency $\omega$ and is 
given by~\citep{sawyer89},
\be
\zeta \approx 6 \times 10^{31} \rho_{15}^2 T_{10}^6 \left ( \frac{\omega}{1\,\mbox{Hz}} \right)^{-2} 
\left [\, 1 + \left (\frac{\omega_\beta}{\omega} \right )^2 \, \right ]^{-1} \mbox{gr}\, \mbox{cm}^{-1}\,\mbox{s}^{-1},
\label{zeta1}
\ee
where $T_{10} = T/10^{10}\,\mbox{K}$, $\rho_{15} = \rho/10^{15}\,\mbox{gr}\, \mbox{cm}^{-3}$ 
and $\omega_\beta$ is the beta reactions characteristic frequency. For the modified Urca process
we have \citep{sawyer89}:
\be
\omega_\beta \approx 0.134\,  \rho_{15}^{-2/3} T_{10}^6\,~\mbox{Hz}.
\label{omb}
\ee
The secondary flow mode frequency $\omega=\omega_{\rm pr}$  can be sufficiently low as to make
$\omega_{\rm pr} \gtrsim \omega_\beta$, causing $\zeta$ to become a non-monotonic function of $T$.
The maximum is attained at the temperature for which $\omega_{\rm pr} = \omega_\beta$. 
This temperature, which also marks the minimum spin-flip timescale, coincides with the adiabatic flow threshold 
$T_{\rm ad}$, see Eqn.~(\ref{Tad}) below.


\subsection{The secondary flow}
\label{app:flow}

The detailed calculation of the secondary flow can be found in~\citet{mestel72}. 
Here we provide an outline of the same calculation albeit with one key difference, namely, we do not 
assume an adiabatic secondary flow.

\citet{mestel72} show quite generally that the hydrostatic equilibrium of a deformed precessing body (i.e. an oblique rotator) requires 
the presence of a small secondary velocity field $\delta \mathbf{v}$ superimposed to the system's bulk rigid body rotation. 
Then, they go on and show that  the pressure and density perturbations $\delta p, \delta \rho$ associated with this secondary flow obey,
\be
\delta p = \frac{dp_0}{d\rho_0} \delta \rho,
\label{dp1}
\ee
where $p_0 (r), \rho_0 (r)$ are the parameters of the non-rotating and non-magnetic spherical star.
With the help of the displacement field $\delta \mathbf{v} = \partial_t\boldsymbol{\xi}$ (this is defined in the stellar rest frame) 
we can write this relation in terms of the Lagrangian pressure and density perturbations. We first have,
\be
\Delta p \approx \delta p + \xi^r p_0^\prime, \qquad \Delta \rho = - \rho_0 \nabla \cdot \boldsymbol{\xi}  \approx  \delta \rho + \xi^r \rho_0^\prime,
\ee
where a prime stands for a radial derivative. Then (\ref{dp1}) becomes,
\be
 \Delta p  = \Gamma \frac{p_0}{\rho_0}  \Delta \rho,
\label{Dp1}
\ee
where we have defined the adiabatic index of the spherical star:
\be
\Gamma \equiv \frac{d\log p_0}{d\log \rho_0}.
\ee

The same perturbations can also be linked  ``microphysically" through the full equation of state. 
Assuming uniform temperature neutron star matter, the equation of state should be of the
bi-parametric form $p=p(\rho,x_\rp)$ where the second degree of freedom (composition) is represented by the proton
fraction $x_\rp$. This implies the following expression for a $\boldsymbol{\xi}$-displaced fluid element:  
\be
\Delta p = \frac{\gamma p}{\rho}  \Delta \rho +  \frac{\partial p}{\partial x_\rp} \Delta x_\rp,
\label{Dp3}
\ee
where we have introduced the adiabatic index for a ``frozen" composition \citep{passamonti09}
\be
\gamma \equiv \left (\frac{d\log p}{d\log \rho} \right )_{x_\rp} = 
\Gamma - \left (\frac{\partial \log p}{\partial \log x_\rp} \right )_{\rho} \frac{d\log x_\rp}{d\log \rho} .
\ee
Combining (\ref{Dp1}) and (\ref{Dp3}) and approximating $p \approx p_0, \rho \approx \rho_0 $, we can arrive at:
\be
 \left ( \Gamma - \gamma \right )  \nabla \cdot \boldsymbol{\xi} =  - \frac{\partial p}{\partial x_\rp} \frac{\Delta x_\rp}{p_0}.
\label{divxi1}
\ee
Given that $\Gamma \neq \gamma$, this equation directly links the compressibility
of the secondary flow and its degree of adiabaticity. An adiabatic flow entails a fixed composition in
a displaced fluid element, i.e. $\Delta x_\rp =0$, which subsequently leads to $\nabla \cdot \boldsymbol{\xi}=0$.
In such a case, the bulk viscosity force is identically zero and the spin-flip instability does not set in~\footnote{It should be 
pointed that although~\citet{mestel72} assume an adiabatic secondary flow, their oblique rotator is dissipative as a result of the 
presence of convection-driven turbulent viscosity in their stellar model.}. 
In the opposite case of a non-adiabatic perturbation we have $\Delta x_\rp \neq 0 \to \nabla \cdot \boldsymbol{\xi} \neq 0 $ 
and therefore the flow couples to bulk viscosity. 

For their assumed adiabatic flow, \citet{mestel72} have $ \nabla \cdot \boldsymbol{\xi}=0$ and obtain the following result for the radial flow:
\bear
&&  \delta v^r  \sim \frac{1}{4} \epsilon_\Omega \epsilon_B \Omega \sin2\chi  \left [\, \sin\chi (1-P_2(\theta) ) 
\sin2(\varphi -\omega_{\rm pr} t)   \right.
\nonumber \\
 && \left. \qquad  + ~ 3 \cos\chi \sin2\theta \sin(\varphi-\omega_{\rm pr} t)  \, \right ] ,
\label{dvr}
\eear
where the angles $\theta,\varphi$ are measured with respect to the magnetic axis and $\chi$ has the same 
meaning as before. This result shows that $\delta \mathbf{v}$ comes in two harmonics and that for a small $\chi$-inclination 
the dominant harmonic is the fundamental one, $\omega =\omega_{\rm pr}$.  
It follows that the amplitude  of the secondary flow scales as,
\be
\delta v \sim  \epsilon_\Omega |\epsilon_B| \Omega R \sin2\chi \cos\chi.
\label{dv_ampl}
\ee
This clearly  shows that $\delta v = 0$ in a state of no-precession, $\chi=\{0,\pi/2\}$. 

Although the velocity (\ref{dvr}) was derived under the assumption of adiabaticity, we should expect the basic scaling
(\ref{dv_ampl}) to hold in general even when the flow is non-adiabatic. We can thus use (\ref{dv_ampl}) for any temperature and
obtain an order of magnitude estimate for the viscous damping (as we indeed have done in Appendix~\ref{app:timescale}).

Given that the composition of a fluid element is controlled by the reaction frequency $\omega_\beta (T)$, the
secondary flow will be non-adiabatic (and viscous) provided $T > T_{\rm ad}$ where the adiabaticity 
threshold $T_{\rm ad}$ is defined as:
\be
 \omega_\beta (T_{\rm ad}) = \omega_{\rm pr} \approx |\epsilon_B| \Omega.
\ee
For the specific case of mURCA reactions we use (\ref{omb}) to obtain,
\be
T_{\rm ad}  \approx 9 \times 10^9\,\rho_{15}^{1/9} P_{\rm ms}^{-1/6} \left ( \frac{\epsilon_B}{10^{-5}} \right )^{1/6}\, ~\mbox{K}.
\label{Tad}
\ee
This is clearly relevant for hot, post-merger neutron stars with strong magnetic fields. 

The same temperature $T_{\rm ad}$ marks the maximum value of the bulk coefficient $\zeta$ and therefore 
corresponds to the minimum spin-flip timescale $\tau_\chi$ (for given deformations $\epsilon_B, \epsilon_\Omega$), see eqn.~(\ref{tau_chi2}).
This can be clearly seen in the curves shown in Figure~\ref{fig:temp}.


\subsection{When does the spin-flip fail?}
\label{app:fail}

According to Figure~\ref{fig:temp} the spin-flip timescale $\tau_{\rm sf}$ and the cooling timescale
typically intersect at a temperature $T_{\rm sf} > T_{\rm ad}$. In such a case the system has the opportunity to spin-flip
and become an orthogonal rotator. However, above some magnetic ellipticity $\epsilon_B = \epsilon_{\rm  sf}$ the two curves will cross 
(if they cross) at a temperature $T < T_{\rm ad}$ and therefore the spin-flip will \textit{not} take place as a consequence of the
secondary flow being incompressible. 

Let us calculate the ellipticity threshold $\epsilon_{\rm sf}$. We need to consider the following two conditions:
\be
T (t_{\rm sf}) = T_{\rm ad}, \qquad  t_{\rm sf} = \tau_{\rm sf} (T_{\rm ad}),
\label{cross1}
\ee
where the mURCA temperature profile is given by eqn.~(\ref{eqn:mURCA}).
From the first condition (\ref{cross1}) we obtain (after omitting the unimportant $T(0)$ term)
\be
t_{\rm sf} \approx 30\, \rho_{15}^{-2/3} P_{\rm ms} \left ( \frac{\epsilon_B}{10^{-5}} \right)^{-1}\,~ \mbox{s}.
\label{tcross}
\ee
Using this result in the second condition (\ref{cross1}) we obtain, after setting $\epsilon_B = \epsilon_{\rm sf} $,
\be
\epsilon_{\rm sf} \approx  10^{-3} \rho_{15} P_{\rm ms}^{-2} R_6^{-2}.
\ee
This ellipticity is represented by the blue bars in Figure~\ref{ellipticity}. 
Inserting the obtained $\epsilon_{\rm sf}$ back into (\ref{tcross}),
\be
t_{\rm sf} \approx 0.3\, \rho_{15}^{-5/3} P_{\rm ms}^3 R_6^2\, \mbox{s}.
\ee
This result is in good agreement with Figure~\ref{fig:temp}  
and implies that, provided $\epsilon_B \gtrsim \epsilon_{\rm sf} $,
the system enters  the `no spin-flip' regime almost instantly. 

The spin-flip instability may be suppressed by an altogether different reason. 
In the high-temperature regime above $10^{10} K$, the neutron star matter becomes opaque to
neutrinos and the previous expressions for the bulk viscosity coefficient and the cooling rate need to be modified
accordingly.
We can estimate the temperature for the transition from neutrino-transparent to opaque matter by using   
the relevant mean free path formula from~\citet{shapiro83}. The result of this exercise is 
$\lambda_\nu \approx 2\times 10^7 (T/10\,\mbox{K})^{-5/2}\,\mbox{cm}$, which implies matter becomes opaque (i.e. $\lambda_\nu < R$)
for $T \gtrsim 3 \times 10^{10}\,\mbox{K}$. 
\citet{lai01} provides a result for the neutrino-opaque bulk viscosity coefficient which
is comparable to (\ref{zeta1}) but with a markedly higher beta reaction frequency 
$\omega_\beta \approx 570\,\rho_{15}^{2/3} T_{10}^4\, \mbox{Hz}$. This modification implies an exceedingly 
low bulk viscosity damping rate (the coefficient $\zeta$ is reduced by several orders of magnitude) which in turn translates 
to a very long spin-flip timescale.


\section*{Acknowledgements}

We are grateful to Antonia Rowlinson, Yuri Levin, Eric Thrane, Andrew Melatos, Eric Howell, and Pantelis Pnigouras for enlightening conversations and comments on the manuscript.  We are also grateful to the anonymous referee for useful feedback that has improved the manuscript.
PDL is supported by an Australian Research Council Discovery Project DP1410102578.  
KG is supported by the Ram\'{o}n y Cajal Programme of the Spanish Ministerio de Ciencia e Innovaci\'{o}n, by a Humboldt Foundation research 
visit grant at the University of T\"ubingen, and also acknowledges travel support from NewCompstar (a COST-funded Research Networking Programme).

\bibliographystyle{mn2e}
\bibliography{PlateauBib}

\label{lastpage}

\end{document}